\documentclass[journal]{IEEETran}  

\IEEEoverridecommandlockouts                              

\usepackage{amsmath}
\usepackage{subfigure}
\usepackage{graphicx}
\usepackage{url}
\usepackage{algorithm}
\usepackage{algorithmic}
\usepackage{amssymb}
\def\xx{{\boldsymbol{x}}}
 \def\B{{B}}
 \DeclareMathOperator*{\rank}{rank}
\def\X{{\boldsymbol{X}}}  
 \def\BB{{\bf B}}
\def\AA{{\boldsymbol{A}}}

\def\Y{{\boldsymbol{Y}}}  

\def\qq{{\boldsymbol{q}}} 
 \def\yy{{\boldsymbol{y}}}
\def\bb{{\boldsymbol{b}}}

\def\Q{{\boldsymbol{Q}}}

\def\subjto{{\mbox{subj. to}}}

\RequirePackage{xspace} \RequirePackage{amsmath, amssymb,color}

\DeclareMathOperator*{\Tr}{Tr}

 \def\xx{{\boldsymbol{x}}}
 \def\B{{B}}
 \def\BB{{\bf B}}
  
 \def\yy{{\boldsymbol{y}}}
\def\bb{{\boldsymbol{b}}}

\def\subjto{{\mbox{subj. to}}}

\renewcommand{\Re}{{\mathbb{R}}}
\newcommand{\T}{\mathsf{T}}
\newcommand{\eg}{\textit{e.g.,~}}
\newcommand{\etc}{\textit{etc~}}
\newcommand{\ie}{\textit{i.e.,~}}
\renewcommand{\H}{\mathsf{H}}

 \newtheorem{thm}{Theorem}
 \newtheorem{ex}{Example}

\newtheorem{df}[thm]{Definition}

\def\Z{{\boldsymbol{Z}}}

\title{\LARGE \bf
Nonlinear Basis Pursuit 
}

\author{Henrik Ohlsson, Allen Y. Yang, Roy Dong and S. Shankar Sastry
\thanks{Ohlsson gratefully acknowledge support by the Swedish Research
  Council in the Linnaeus center CADICS, the European Research Council
   under the advanced grant LEARN, contract 267381, by a postdoctoral grant from the Sweden-America
   Foundation, donated by ASEA's Fellowship Fund, and  by a postdoctoral
   grant from the Swedish Research Council. Yang is supported in part by ARO 63092-MA-II. Yang and Sastry are also supported in part by DARPA FA8650-11-1-7153 and ONR N00014-12-1-0609.}
\thanks{Ohlsson, Yang, Dong, and Sastry are with the Department of Electrical Engineering and Computer  Sciences, University of California, Berkeley, CA, USA. Ohlsson is also with the
Division of Automatic Control, Department of Electrical Engineering, Link\"oping University, Sweden.
{\tt\small  ohlsson@eecs.berkeley.edu.}}}%

\begin{document}

\maketitle
\thispagestyle{empty}
\pagestyle{empty}

\begin{abstract}
In compressive sensing, the basis pursuit algorithm aims to find the sparsest solution to an underdetermined
 linear  equation system. In this paper, we generalize basis pursuit to finding the sparsest solution to
 higher order nonlinear systems of equations, called nonlinear basis pursuit. In contrast to the existing
 nonlinear compressive sensing methods, the new algorithm that solves the nonlinear basis pursuit problem
 is convex and not greedy. The novel algorithm enables the compressive sensing approach to be used for a broader range
 of applications where there are nonlinear relationships between the measurements and the unknowns.

\end{abstract}

\section{Introduction}
Consider the problem of finding the sparsest solution $\xx$ to a linear system of equations:
\begin{equation}\label{eq:spars}
y_i= \bb_i^\T \xx \in \Re, \quad i=1,\dots, N, \quad \xx\in \Re^n,
\end{equation}
where $y_i$ and $\bb_i$ are presumed given.
This problem has received considerable attention recently in the area of compressive sensing.
The exact solution to \eqref{eq:spars} is known as $\ell_0$-minimization ($\ell_0$-min):
\begin{subequations}\label{eq:spars2}
\begin{align}
\min_{\xx}  &\quad \|\xx\|_0 \\  \subjto
& \quad  y_i= \bb_i^\T \xx, \quad i=1,\dots, N.
\end{align} \end{subequations}
It is well known that solving $\ell_0$-min is an NP-hard problem. Hence, it is not practical to directly solve the problem except when the system dimension $n$ is very small. 

In the literature, algorithms for finding an approximate solution to \eqref{eq:spars2} can be
divided into two categories. The first category includes \textit{greedy algorithms} \cite{DavisG1997,NeedellD2008,DaiW2009-TIT}. Because these greedy algorithms iteratively update
their estimate of $\xx$ based on local information, their computational complexity in the processing of computing the estimation updates is significantly
lower than that of $\ell_0$-min. However, their main disadvantage is a weaker guarantee of convergence, especially comparing to the second category below.

The second category of algorithms solves a convex relaxed problem known as $\ell_1$-minimization or \emph{basis pursuit} \cite{Chen:98}:
\begin{subequations}\label{eq:bp}
\begin{align}
\min_{\xx}  &\quad \|\xx\|_1 \\  \subjto
& \quad  y_i= \bb_i^\T \xx, \quad i=1,\dots, N.
\end{align} \end{subequations}
Compared to greedy algorithms, basis pursuit provably recovers the exact solution as $\ell_0$-min under some mild conditions as described in compressive sensing theory \cite{Donoho:03b,CandesConf:06,bruckstein:09}.
As the nonconvex $\ell_0$-norm function is replaced by a convex
$\ell_1$-norm function, basis pursuit can be solved by convex
optimization. In sparse optimization literature, there have been
extensive discussions about accelerating the implementation of basis
pursuit. The interested reader is referred to
\cite{LorisI2009,YangA2013-TIP}. For further readings on greedy
algorithms, basis pursuit, and compressive sensing, see \cite{Eldar:2012}.

More recently, compressive sensing theory has been extended to solving nonlinear problem, which is called nonlinear compressive sensing (NLCS) \cite{BlumensathT:2012,Beck:2012}:
\begin{equation}
\begin{aligned}
\min_{\xx}  &\quad \|\xx\|_0 \\  \subjto
& \quad  y_i= f_i(\xx) \in \Re, \quad i=1,\dots, N,
\end{aligned} 
\label{eq:nlcs}
\end{equation}
where $f_i(\xx)$ in general can be considered a smooth function.

We are particularly interested in solving \eqref{eq:nlcs}, as it can extend the compressive sensing approach to a broad range of applications
where the relationship between the measurements and the unknowns is nonlinear. For example, the problem of \emph{quadratic basis pursuit} (QBP) \cite{Ohlsson:13,ohlsson:11m,Ohlsson:12} is a special case of the NLCS formulation, which is a fundamental solution to the compressive phase retrieval problem in the applications of diffraction imaging \cite{MoravecM2007} and sub-wavelength imaging \cite{Shechtman:11}.

Similar to solving $\ell_0$-min, directly solving \eqref{eq:nlcs} in its original form is very expensive and intractable in practice. Therefore, there is a need
to seek more efficient numerical algorithms to estimate sparse solutions in \eqref{eq:nlcs} while their convergence can be also guaranteed. As the topic
of nonlinear compressive sensing is rather new, there are just a few algorithms that have been proposed in the literature. For instance, one of the first papers discussing this problem is \cite{BlumensathT2008}, which proposed a greedy gradient-based algorithm. Another greedy approach was also proposed in \cite{LiL2009}. In \cite{Beck:2012}, the authors proposed several iterative hard-thresholding and sparse simplex pursuit algorithms. As these algorithms are nonconvex greedy solutions, the analysis of their convergence typically only concerns about their local behavior. In \cite{BlumensathT:2012}, the author also considered a generalization of the restricted isometry property (RIP) to support the use of similar iterative hard-thresholding algorithms for solving general NLCS problems.

In this paper, we present a novel solution to NLCS, called
\emph{nonlinear basis pursuit} (NLBP). The work extends our previous
publications in quadratic compressive sensing and compressive phase
retrieval \cite{ohlsson:11m,Ohlsson:13}, and proposes a convex
algorithm to solve NLBP via a high-order Taylor expansion and a
lifting technique. 
The work was inspired by several recent works on CS applied to the phase retrieval
problem
\cite{MoravecM2007,Marchesin:09,Chai:10,Shechtman:11,Candes:11,Candes:11b,Osherovich:12,Szameit:12,Jaganathan12,schniter12,Shechtman13}. 

The
paper is organized as follows.
We will first discuss some basic notation for this paper in Section \ref{sec:notation}. Then we will present the NLBP problem and its analysis of theoretical convergence in Section \ref{sec:NLBP}. In Section \ref{sec:solver}, we will present a numerical efficient convex solver to estimate the sparse solution of NLBP. Finally, in Section \ref{sec:evaluation}, we will present numerical evaluation to validate the performance of our algorithm and compare with other existing solutions.

\section{Notation and Assumptions}
\label{sec:notation}
In this paper, we use bold face to denote vectors and matrices, and normal font
for scalars. 
We denote
the transpose of a real vector by  $\xx^\T$ and the conjugate
transpose of a complex vector by $\xx^\H$.  $\X({i,j})$ denotes the
$(i,j)$th element.
Similarly, we
let $\xx(i)$ be the $i$th element of the vector $\xx$.
Given two matrices $\X$ and $\Y$,
we use the following fact that their product in the trace function commutes, namely, $\Tr(\X \Y) = \Tr(\Y \X)$, under the assumption
that the dimensions match. $\| \cdot \|_0$ counts the number of nonzero elements in a vector or matrix; similarly, \linebreak $\| \cdot\|_1$ denotes the element-wise $\ell_1$-norm of a vector or matrix, \ie, the sum of the magnitudes of the elements; whereas $\| \cdot \|$ represents the $\ell_2$-norm for vectors and the spectral norm for matrices.
 We assume that the functions $f_i(\cdot),i=1,\dots,N$ are analytic functions. 
 
\section{Nonlinear Basis Pursuit}
\label{sec:NLBP}
Since $f_i(\cdot),i=1,\dots,N$, are analytic functions, we can express
them using their Taylor expansions.  
Using \textit{multi-index notation}, we can write the Taylor expansion of $f_i$
around $\xx_0$ as 
\begin{equation}
f_i(\xx_0)=\sum_{|\alpha|\geq 0}  \frac{(\xx - \xx_0)^\alpha}{\alpha!}
\partial^\alpha f_i(\xx_0).
\end{equation}
If $q$ is an even integer, we can now rewrite a $q$ order Taylor
expansion as
\begin{align}
f_i(\xx_0)=&\sum_{0\le|\alpha|\le q}  \frac{(\xx - \xx_0)^\alpha}{\alpha!}
\partial^\alpha f_i(\xx_0) \\ \label{eq:quadratic} =& 
\bar \xx^\T \Q_i \bar \xx 
\end{align}
where $\Q_i$ is a ${n+ \frac{q}{2}\choose \frac{q}{2}}\times {n+\frac{q}{2}\choose \frac{q}{2}}$-symmetric matrix, and $\bar \xx$ contains all the monomials of the elements of $\xx$ with degree
$\leq q/2$.
\begin{ex}\label{ex:firstc}
Let $\xx=[x_1 \,x_2]^T$ and $f(\xx)=1+x_1+x_2^3$. We consider a 4th order Taylor expansion
around $\xx_0 =0$. Then the set of 2-tuple $\alpha$'s is 
\begin{equation}
\begin{tabular}{rcl}
A&=&\{(0,0), (1,0), (0,1), (1,1), (2,0), (0,2), (2,1), (1,2),\\
 & &(3,0), (0,3), (2,2), (3,1), (1,3), (4,0), (0,4)\}.
\end{tabular}
\end{equation}
We can easily verify that $|A | = {n+q \choose q} = {2+4 \choose 4} = 15$.

Hence, we can rewrite $f(\xx)$ as
\begin{equation*}
f(\xx)=
\begin{bmatrix} 1 \\ x_1\\ x_2 \\ x_1x_2 \\ x_1^2 \\
  x_2^2 \end{bmatrix}^\T\begin{bmatrix} 1 & 1/2&0&0&0&0 \\
  1/2&0&0&0&1/12&0\\  0&0&0&0&0&0 \\  0&0&0&0&0&0 \\  0&1/12&0&0&0&0 \\  0&0&0&0&0&0 \end{bmatrix}\begin{bmatrix} 1 \\ x_1\\ x_2 \\ x_1x_2 \\ x_1^2 \\
  x_2^2 \end{bmatrix}.
\end{equation*}
Again, we verify that the dimension of $\bar{\xx}$ is equal to ${2+2 \choose 2} = 6$.
\end{ex}

From the example it is easy to see that elements of $\bar \xx$ are
generally  dependent. 
The dependencies between elements of $\bar \xx$
needs to be made explicit. 

\begin{ex}
Let, as in the previous example  
\begin{equation} \bar \xx =\begin{bmatrix} 1 & x_1 & x_2 & x_1x_2 & x_1^2 &
  x_2^2 \end{bmatrix}^\T.
\end{equation}
It is clear that \eg the second element $x_1$ times the third  element
$x_2$ gives the third element  $x_1x_2$ of $\bar \xx$. This can be
expressed as
\begin{equation*}
0=
\begin{bmatrix} 1 \\ x_1\\ x_2 \\ x_1x_2 \\ x_1^2 \\
  x_2^2 \end{bmatrix}^\T\begin{bmatrix} 0 & 0&0&-1/2&0&0 \\
  0&0&1/2&0&0&0\\  0&1/2&0&0&0&0 \\  -1/2&0&0&0&0&0 \\  0&0&0&0&0&0 \\  0&0&0&0&0&0 \end{bmatrix}\begin{bmatrix} 1 \\ x_1\\ x_2 \\ x_1x_2 \\ x_1^2 \\
  x_2^2 \end{bmatrix}.
\end{equation*}
\end{ex}

Constructively, the dependences between elements can be generated as
follows.  Let $\Q_{N+1}$ be a  ${n+
    \frac{q}{2}\choose \frac{q}{2}}\times {n+\frac{q}{2}\choose
    \frac{q}{2}}$-symmetric matrix with $\Q_{N+1}(1,1)=1$ and set all
  other elements to zero. Set $y_{N+1}=1$, $i=1,$ $k=l=2$ and $m=N+2$.
\begin{enumerate}
\item If $k\neq l$ and $\bar  \xx(i) =  \bar  \xx(k) \bar
  \xx(l)$,    let $\Q_m$ be a  ${n+
    \frac{q}{2}\choose \frac{q}{2}}\times {n+\frac{q}{2}\choose
    \frac{q}{2}}$-symmetric matrix with $\Q_m(k,l)=\Q_m(l,k)=1/2$,
  $\Q_m(1,i)=\Q_m(i,1)=-1/2$, set all other elements to zero, $y_m=0$ and set
  $m=m+1$. 

\item If $k= l$ and $\bar  \xx(i) =  \bar  \xx(k) \bar
  \xx(l)$,    let $\Q_m$ be a  ${n+
    \frac{q}{2}\choose \frac{q}{2}}\times {n+\frac{q}{2}\choose
    \frac{q}{2}}$-symmetric matrix with $\Q_m(l,l)=1$,
  $\Q_m(1,i)=\Q_m(i,1)=-1/2$, set all other elements to zero, $y_m=0$ and set
  $m=m+1$. 

\item If $k <  {n+
    \frac{q}{2}\choose \frac{q}{2}} $, set $k=k+1$ and return to step
  1, otherwise continue to the next step.
\item If $l <  {n+
    \frac{q}{2}\choose \frac{q}{2}} $, set $l=l+1$, $k=2$ and return to step
  1, otherwise continue to the next step.
\item If $i <  {n+
    \frac{q}{2}\choose \frac{q}{2}} $,  set $i=i+1$, $k=l=2$ and
  return to step 1. Otherwise set $M=m-1$, return $\{(y_i,\Q_i)\}_{i=1}^M$ and abort.
\end{enumerate}
The dependencies can now be expressed as 
\begin{align}
y_i=&\bar \xx \Q_i \bar \xx,\quad i=N+1,\dots,M.
\end{align}
Note that these constraints are all quadratic. 

Assuming that the $q$th order Taylor expansion is a good
approximation for $\{f_i(\xx)\}_{i=1}^N$, then the problem of finding the sparsest
solution to \eqref{eq:nlcs} can be written as:
\begin{equation}
\begin{aligned}
\min_{\bar \xx}  &\quad \|\bar \xx\|_0 \\ 
\subjto & \quad  y_i= \bar \xx^\T  \Q_i \bar \xx, \quad
i=1,\dots, M.
\end{aligned}
\label{eq:spars4}
\end{equation}
Next, we employ a lifting technique used extensively in quadratic
programming \cite{shor87,Lovasz91,Nesterov98,Goemans:1995} and define
a positive semi-definite matrix $\bar \xx \bar \xx^\T=\X$, which
satisfies, for $i=1,\dots,M$, the following equalities
\begin{equation}
y_i= \bar
\xx^\T  \Q_i \bar \xx =\Tr (\bar
\xx^\T \Q_i \bar \xx)=\Tr ( \Q_i \bar \xx \bar
\xx^\T)= \Tr ( \Q_i 
\X). 
\end{equation}
The nonconvex problem in \eqref{eq:spars4} can now be shown equivalent to 
\begin{equation}
\begin{aligned}
\min_{\X}  &\quad \|\X\|_0 \\ 
\subjto
& \quad y_i= \Tr (  \Q_i \X), \quad i=1,\dots, M, \\ & \quad  \rank(\X) =1, \:\X\succeq 0.
\end{aligned} 
\label{eq:spars5}
\end{equation}

Due to the $\ell_0$-norm function and the rank condition in \eqref{eq:spars5},
the problem is still combinatorial. To relax the $\ell_0$-norm
function in \eqref{eq:spars5}, we can replace the $\ell_0$-norm with
the $\ell_1$-norm.  
To relax the rank condition, we can remove the rank constraint and instead minimize $\rank(\X)$ in the objective function. Furthermore, since the rank
of a matrix is still not a convex function, we choose the replace the rank with the nuclear norm of $\X$, which is known to be a convex heuristic of
the rank function \cite{Fazel01arank,CANDES:2009}. For a semidefinite
matrix $\X$, it is also equal to the trace of $\X$. Note that there
are several other heuristics for the rank. For example, the $\log\det$
heuristic described in \cite{Fazel01arank} is an interesting
alternative (see also \cite{Shechtman:11}). However, we choose to use the nuclear norm heuristic here
even
though our theoretical results also holds for \eg the
$\log\det$ heuristic.   
We finally  obtain the convex program  
\begin{equation}\label{eq:spars6}
\begin{aligned}
\min_{\X}  & \quad  \Tr(\X)+\lambda\|\X\|_1 \\ \subjto
& \quad y_i= \Tr ( \Q_i \X), \quad i=1,\dots, M, \\ & \quad  \X\succeq 0,
\end{aligned} \end{equation} 
where $\lambda \geq 0$ is a design variable. We refer to \eqref{eq:spars6} as \emph{nonlinear basis pursuit} (NLBP). 

\section{Theoretical Analysis}
NLBP is a convex relaxation of the combinatorial
problem given in
\eqref{eq:spars4}. It is of interest to know when the relaxation is
tight. As a special case, when the degree of the analytic functions $f_i(\xx)$ are no greater than two, 
the problem is known as \emph{quadratic basis pursuit} (QBP) \cite{Ohlsson:13,ohlsson:11m}.
In fact, the underlying optimization problem of NLBP is the same as that of QBP. Therefore, the theoretical
analysis about the performance guarantees of QBP also applies to NLBP in this paper. In this section, we
only highlight several key results that extend from the proofs given in \cite{Ohlsson:13,ohlsson:11m}.

First, it is convenient to introduce a linear operator $\B$:
\begin{equation}
\B: \X\in \Re^{{n+ \frac{q}{2}\choose \frac{q}{2}}\times {n+ \frac{q}{2}\choose \frac{q}{2}}} \mapsto \{\Tr (\Q_i \X) \}_{1\le i \le M}\in\Re^{M}.
\label{eq:definition-B}
\end{equation}
 We consider a generalization of the \emph{restricted isometry property} (RIP) of the linear operator $\B$. 
\begin{df}[\bf RIP]\label{def:RIP}
A linear operator $\B(\cdot) $ as defined in \eqref{eq:definition-B} is $(\epsilon,
k)$-RIP 
if 
\begin{equation}\label{eq:RIP}
\left |{ \frac{\| \B(\X) \|^2}{\| \X \|^2}} -1
  \right |<\epsilon
\end{equation} for all $\|\X\|_0 \leq k$ and $\X\neq 0.$
\end{df}

We can now state the following theorem:
\begin{thm}[\bf Guaranteed recovery using RIP]\label{thm:guartee1}
Let $\bar \xx$ be the  solution to \eqref{eq:spars4}. 
 The solution of NLBP $\tilde \X$ is equal to  $\bar
  \xx \bar \xx^\T $ if  it
has rank 1 and
$\B(\cdot) $ is ($\epsilon, 2\|\tilde \X\|_0$)-RIP with $\epsilon<1$.
\end{thm}

The RIP-type argument may be difficult to check for a given matrix and
are more useful for claiming results for classes of matrices/linear
operators. For instance, it has  been shown that random Gaussian matrices satisfy the RIP  with
high probability. However, given realization of a random Gaussian matrix,
it is indeed difficult to check if it actually satisfies the RIP.

In the literature, there exist two alternative arguments, namely, the \emph{spark condition} \cite{Chen:98} and the \emph{mutual coherence} \cite{Donoho:03b}.
The spark condition usually gives tighter bounds but is known to be difficult to
compute as well. On the other hand, mutual coherence may give less tight bounds,
but is more tractable. Next, we focus our discussion on mutual coherence, which is defined as: 
\begin{df}[\bf Mutual coherence]
 For a matrix $\AA$,  define  the \textit{mutual coherence}  as
 \begin{equation}
 \mu(\AA)=\max_{1\leq i,j \leq n, i\neq j} {  \frac{|\AA_{:,i}^\T
  \AA_{:,j} |}{\|\AA_{:,i}\|\|\AA_{:,j}\|}}.
  \end{equation}
\end{df}

Let $\BB$ be the matrix satisfying $\yy= \BB \X^s=\B(\X)$ with $\X^s$ being the vectorized version of $\X$. We are now ready to state the following theorem:
\begin{thm}[\bf Recovery using mutual coherence]\label{thm:guartee2}
Let $\bar \xx$ be the solution to \eqref{eq:spars4}. 
 The
solution of NLBP, $\tilde \X$, is equal to  $\bar
  \xx \bar \xx^\T $ if  it
has rank 1 and
$\|\tilde \X\|_0 < 0.5(1+1/\mu(\BB )).$
\end{thm}

\section{Numerical Solvers}
\label{sec:solver}

Naturally, efficient numerical solvers that implement nonlinear basis
pursuit are desirable. There are many classes of methods, commonly
used in compressed sensing, which can be used to solve non-smooth
SDPs. Among these include interior point methods, which is used in the
popular software package CVX \cite{cvx1}, gradient projection methods
\cite{BertsekasD1999}, and augmented Lagrangian methods (ALM)
\cite{BertsekasD1999}, outer approximation methods \cite{Konno2002},
and the alternating directions method of multipliers (ADMM), see for
instance \cite{BertsekasParallel,boyd:11}.

Interior point methods are known generally to scale poorly to moderate- or large-scale problems. Gradient projection methods require a projection onto the feasible set; for our problem, this feasible set is the intersection of a subspace with the positive semidefinite cone. The complexity of this projection operator limits the benefits of using a gradient projection method. For ALM, the augmented primal and dual objection functions are still SDPs, which are equally difficult to solve in each iteration as the original problem. Also, we found that outer approximation methods converge very slowly.

On the other hand, \eqref{eq:spars6} decomposes nicely into the ADMM framework. That is, we can define the following functions:
\begin{subequations}
\begin{align}
h_1(\X) & =
	\begin{cases}
		\Tr(\X) & \mbox{if }  y_i = \Tr( \Q_i \X), \quad i=1,\dots, M \\
		\infty & \mbox{otherwise}
	\end{cases}
\\
h_2(\X) & =
	\begin{cases}
		0 & \mbox{if } \X \succeq 0 \\
		\infty & \mbox{otherwise}
	\end{cases}
\\
g(\Z) & = \lambda \|\Z\|_1
\end{align}
\end{subequations}
Then, \eqref{eq:spars6} becomes:
\begin{equation}
\begin{aligned}
\min_{\X_1,\X_2,\Z} & \quad  h_1(\X_1) + h_2(\X_2) + g(\Z) \\
\subjto 			& \quad \X_1 = \X_2 = \Z
\end{aligned}
\end{equation}
which has the form that ADMM applies to.
ADMM has strong convergence results, converges quickly in practice, and scales well to large data sets \cite{boyd:11}. 
For more details of this implementation, we refer the reader to \cite{Ohlsson:13,ohlsson:11m}.


\section{Numerical Evaluation}
\label{sec:evaluation}
\subsection{A Simple Example}
Let  $n=5$, $N=50$ and consider 
\begin{equation}
y_i=\sum_{4\geq |\alpha|\geq 0}  \qq_{\alpha} \xx ^\alpha,\quad
i=1,\dots, N,\;\xx\in \Re^n, 
\end{equation}
$ \{\qq_{\alpha}\}_ {4\geq |\alpha|\geq 0}$ was generate by sampling
from a Gaussian distribution and $\xx$ was zero except for two
elements that both were one. A mote Carlo simulation consisting of 100
trials was performed. We compared NLBP (fourth order Taylor expansion), QBP (second
order Taylor expansion), and LASSO (first order Taylor
expansion). $\xx_0=0$ was used. The results
are given in Table~\ref{tab:MCMC}.
 \begin{table}[h!]
\caption{Success rates of NLBP, QBP and LASSO for $N=50$ and $n=5$.}
\label{tab:MCMC}
\begin{center}
\begin{tabular}{l|ccc}
\text{\bf Method}& NLBP&QBP & LASSO
\\ \hline 
     Success rates     &100\% & 74\%&0\%\\
\end{tabular} 
\end{center}
\end{table}

\subsection{Polynomial Equation System}
\begin{figure}[h!]
\centering
\includegraphics[width=0.8\columnwidth]{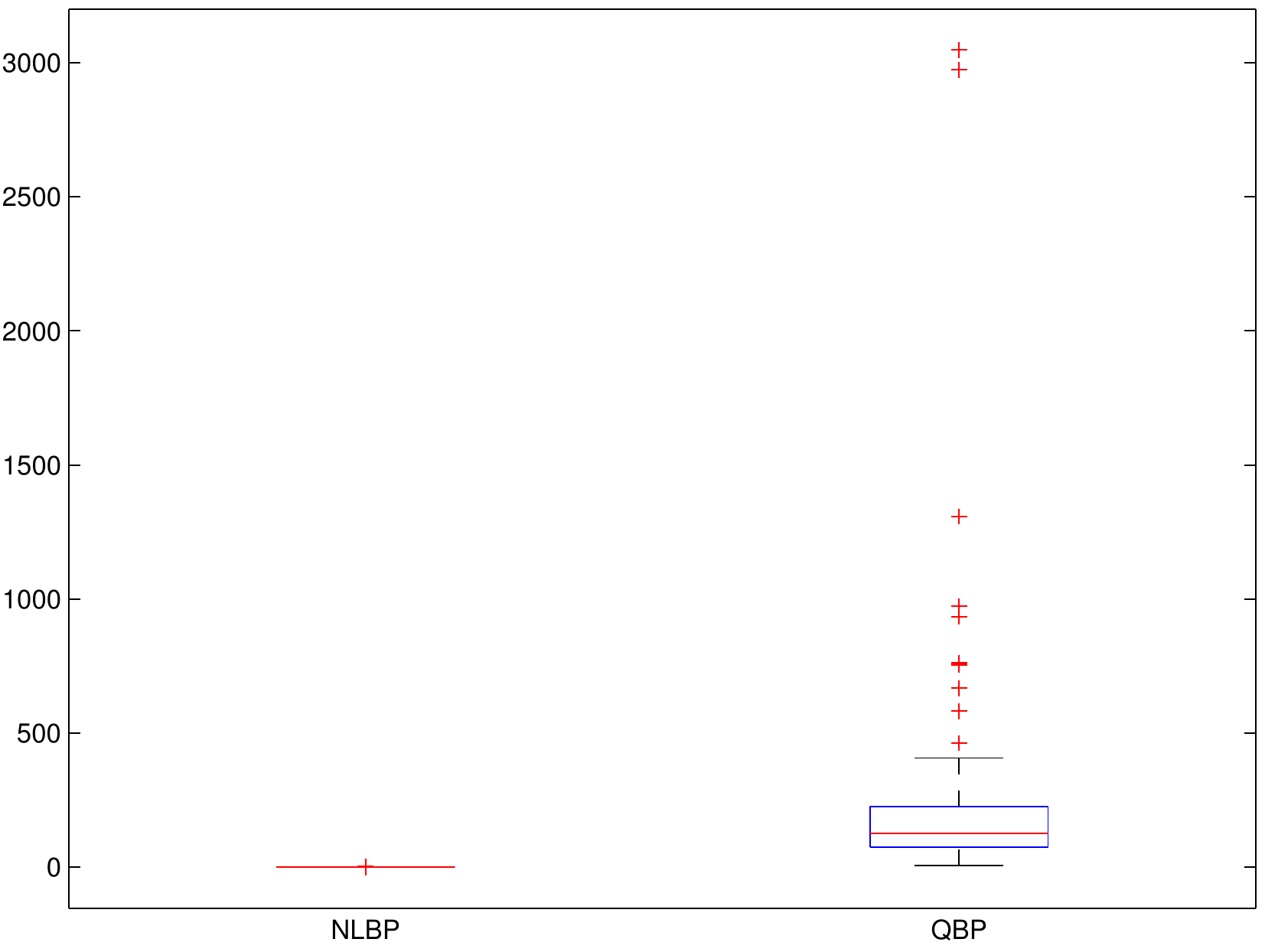}
\caption{Box plot comparing NLBP and QBP for finding the dense
  solution of polynomial equation systems. The box plot summarizes the
result from 100 trials.}\label{fig:box}
\end{figure}

In this example we consider the problem of finding the dense $\xx$ that
solves a system of 4th order polynomials
\begin{equation}
y_i=\sum_{4\geq |\alpha|\geq 0}  \qq_{\alpha} \xx ^\alpha,\quad
i=1,\dots, N,\;\xx\in \Re^n, 
\end{equation}
given the coefficients $ \{\qq_{\alpha}\}_ {4\geq |\alpha|\geq 0}$.
We will let $N=60$, $n=5$ and generate $\xx$ by sampling from a Gaussian
distribution with a standard deviation of 10. Figure \ref{fig:box}
shows the box plots of the squared residuals from a monte Carlo simulation consisting of 100
trials. New polynomial coefficients were generated at random for
each trial and a new  $\xx$. NLBP found the true solution
(within machine precision) in 99
out of the 100 trials. QBP never found the correct solution.   Both
used $\lambda=0$. LASSO with $\lambda=0$ (the least squares estimate) did not give
a satisfactory estimate.
\section{Conclusion and Future Work}

The main contribution of this paper is a nonlinear compressive sensing
algorithms based on convex relaxations. The algorithm, referred to as
nonlinear  basis pursuit,  is rather general
in   that it applies to any analytic nonlinearity by approximate it by
a  Taylor expansion of desired order. Nonlinear basis pursuit inherits
theoretical guarantees, such as guaranteed recovery \etc from its
linear relative (basis pursuit) and therefore  comes with
theoretical guarantees that greedy algorithms often lack. Nonlinear
basis pursuit  takes the form of a convex non-smooth
SDP which can be solved using conventional software for problems of
interest. 

It should be noticed that solving a nonlinear equation system is by itself a
difficult problem. It is therefore quiet remarkable that we with
rather high succes rate manage to find the sparsest solution to the
nonlinear equation system. In addition, solving an overdetermined
nonlinear equation system is also difficult. As shown in the example
section, nonlinear basis pursuit not only find sparse solutions but
also can also provide dense solutions to nonlinear equation system when no sparse solutions is available.

Convexifying nonlinear contraints using SDPs via its Taylor expansion is up
to our knowledge novel and should find applications in many
areas. This is seen as future research.  Last, we have not discussed noise in this paper. However, this extension is
trivial and a nonlinear extension of basis pursuit denoising follows. 

\section{ACKNOWLEDGMENTS}
The authors would like to acknowledge useful discussions and inputs from Yonina C. Eldar, Laura Waller, Filipe Maia, Stefano Marchesini and Michael Lustig.


\bibliographystyle{plain}
\bibliography{refHO}
\end{document}